\documentclass[letterpaper, 10 pt, conference]{IEEEtran}
\IEEEoverridecommandlockouts
\usepackage{cite}
\usepackage{amsmath,amssymb,amsfonts}
\usepackage{algorithmic}
\usepackage{graphicx}
\usepackage{textcomp}
\usepackage[dvipsnames]{xcolor}
\usepackage{physics}
\usepackage{amssymb}

\newtheorem{definition}{Definition}[section]

\def\BibTeX{{\rm B\kern-.05em{\sc i\kern-.025em b}\kern-.08em
    T\kern-.1667em\lower.7ex\hbox{E}\kern-.125emX}}
\begin{document}

\title{Weaving Complex Graph on simple low-dimensional qubit lattices}

\author{\IEEEauthorblockN{1\textsuperscript{st} Yu-Hang Dang*}
\IEEEauthorblockA{\textit{Shenzhen Institute for Quantum Science and Engineering} \\
\textit{Southern University of Science and Technology}\\
Shenzhen, P. R. China}
\and
\IEEEauthorblockN{2\textsuperscript{nd} Shyam Dhamapurkar*}
\IEEEauthorblockA{\textit{Shenzhen Institute for Quantum Science and Engineering} \\
\textit{Southern University of Science and Technology}\\
Shenzhen, P. R. China}
\and
\IEEEauthorblockN{3\textsuperscript{rd} Xiao-Long Zhu}
\IEEEauthorblockA{\textit{Shenzhen Institute for Quantum Science and Engineering} \\
\textit{Southern University of Science and Technology}\\
Shenzhen, P. R. China}
\and
\IEEEauthorblockN{4\textsuperscript{th} Zheng-Yang Zhou}
\IEEEauthorblockA{\textit{Southern University of Science and Technology}\\
Shenzhen, P. R. China}
\and
\IEEEauthorblockN{5\textsuperscript{th} Hao-Yu Guan}
\IEEEauthorblockA{\textit{Shenzhen Institute for Quantum Science and Engineering} \\
\textit{Southern University of Science and Technology}\\
Shenzhen, P. R. China}
\and
\IEEEauthorblockN{6\textsuperscript{th} Xiu-Hao Deng}
\IEEEauthorblockA{\textit{Shenzhen Institute for Quantum Science and Engineering} \\
\textit{Southern University of Science and Technology}\\
Shenzhen, P. R. China\\}
\IEEEauthorblockA{\textit{International Quantum Academy}\\
Shenzhen, P. R. China\\
dengxh@sustech.edu.cn}
}

\maketitle

\begin{abstract}
In quantum computing, the connectivity of qubits placed on two-dimensional chips limits the scalability and functionality of solid-state quantum computers. This paper presents two approaches to constructing complex quantum networks from simple qubit arrays, specifically grid lattices. The first approach utilizes a subset of qubits as tunable couplers, effectively yielding a range of non-trivial graph-based Hamiltonians. The second approach employs dynamic graph engineering by periodically activating and deactivating couplers, enabling the creation of effective quantum walks with longer-range couplings. Numerical simulations verify the effective dynamics of these approaches. In terms of these two approaches, we explore implementing various graphs, including cubes and fullerenes, etc, on two-dimensional lattices. These techniques facilitate the realization of analog quantum simulation, particularly continuous-time quantum walks discussed in detail in this manuscript, for different computational tasks on superconducting quantum chips despite their inherent low dimensional simple architecture. \end{abstract}

\begin{IEEEkeywords}
Effective Hamiltonian, Superconducting qubits, Dynamic graphs, Quantum walks 
\end{IEEEkeywords}

\def\thefootnote{}\footnotetext{*These authors contributed equally to this work.}

\section{Introduction}
With the rapid development of quantum computing hardware, particularly superconducting (SC) qubit systems~\cite{blais2020quantum}, the ability to implement complex quantum algorithms and simulations is critical for leveraging the power of these devices. However, a key challenge arises from the limited connectivity imposed by fabricating qubits on one- or two-dimensional (1D or 2D) arrays. This inherent constraint on the qubit coupling architecture limits the scalability and functionality achievable with quantum computers~\cite{cheng2023noisy,burkard2023semiconductor,wintersperger2023neutral}.

One of the most essential quantum tasks where qubit connectivity is crucial is analog quantum simulation. Particularly, we use continuous-time quantum walks (CTQWs) as an example to illustrate how to improve effective connectivity. CTQWs, the quantum equivalent of classical random walks, involve walkers transitioning between vertices (qubits) on a graph while in a quantum superposition. Based on CTQWs on various graphs, quantum algorithms are being proposed with significant computational speedups over their classical counterparts ~\cite{Magniez05quantumalgorithms, Ambainis_element_distinctness, Childs_2004, childs2003exponential, richter2007quantum, childs2003quantum}. Demonstrations of such speedups have been studied extensively on integrated photonic systems with the lack of graph tunability~\cite{tang2018experimental}. In recent years, there have been some experiments implementing CTQWs on controllable multi-qubit systems ~\cite{qiang2021implementing, gong2021quantum, young2022tweezer}. However, realizing arbitrary graphs required for different CTQW applications poses a formidable challenge on 1D or 2D qubit arrays with nearest-neighbour couplings.

In this work, we introduce two novel methods that facilitate the construction of intricate quantum networks and graphs on simple, low-dimensional qubit lattices found in SC chips. 1. \textit{Static edge weaving (SEW)}: This technique statically detunes a subset of qubits as effective "weaving" edges that bridge distant vertices on the lattice. 2. \textit{Periodic edge weaving (PEW)}: This method weaves the graph dynamically by periodically switching qubit couplings on and off to enable the extended-range bridges. Together, these methods allow for the formation of various graph structures by weaving simple low-dimensional lattices. 

Through comprehensive numerical simulations of the complete Hamiltonian, we have verified the effectiveness of these methods in replicating the desired quantum dynamics across various graph structures. Notably, we have successfully executed complex graphs such as complete graphs, glued tetrahedrons, cubes, and fullerenes—structures that are typically challenging to implement directly on planar qubit grids using conventional approaches.

These proposed methods significantly advance analog quantum simulations, focusing on Continuous-Time Quantum Walks (CTQWs) for diverse computational tasks on state-of-the-art SC quantum processors. By addressing the connectivity constraints of low-dimensional qubit architectures, our work paves the way for fully utilizing the capabilities provided by these advanced quantum hardware platforms.

\section{Preliminaries}

Before going into the proposed scenarios and methods, we give some important definitions and clarify some fundamental concepts and terminology.

\subsection{Quantum walks}
Let us define a CTQW on a graph~\cite{chakraborty2020fast}. A CTQW on a graph $G$ with $N$ nodes labelled by $j=1,2,...N$. The Hilbert space for the quantum walk is of dimension $N$, with basis states ${\ket{j}}$ corresponding to each vertex $j$ in $G$. The state of the system at time $t$ is described by the probability amplitudes $\alpha_j(t)$, where $\ket{\psi(t)} = \sum_{j} \alpha_j(t) \ket{j} = e^{-iHt} \ket{\psi(0)}$, where $\ket{\psi(0)} = \ket{j}$. The Hamiltonian $H$ is commonly chosen as the adjacency matrix $A$ of the graph, where

\begin{equation}\label{Eq:adjacency}
    \bra{k}A\ket{j} = \begin{cases}
       1 & j \neq k,  \text{if edge }(j,k) \in G \\
       0     & \text{otherwise}.
    \end{cases}
\end{equation}

If the walk starts at some vertex $j$ and runs for time $t$, then the probability to measure the walker at some vertex $k$ is $P(j,k) = |\bra{k} e^{-iHt}\ket{j}|^2$.    

\subsection{Analog Simulation of Graph-Based CTQW}

CTQW can be simulated with the Bose-Hubbard model (BHM) Hamiltonian \cite{roushan2017spectroscopic} on qubit lattices such as SC qubits ~\cite{gong2021quantum,deng2016superconducting}:
\begin{equation}\label{BHM}
H=\sum_j (\omega_j a^\dagger_j a_j + U_j a^\dagger_j a^\dagger_j a_j a_j)+\sum_{\langle j,k \rangle} g_{jk} (a^\dagger_j a_k + a_j a^\dagger_k),
\end{equation}
where $a_j,a_j^{\dagger}$ are the annihilation and creation operators, respectively; $\omega_j$ are the chemical potential (qubit frequency), $U_j$ are the on-site interaction energy (qubit anharmonicity), and $g_{jk}$ the hopping strength between the site $j$ and $k$. $g_{jk}$ can be tuned in the range of $[0,g_{max}]$ controlled by a tunable coupler~\cite{yan2018tunable}. Here, $J$ is a constant that quantifies the walk speed. The bosons of BHM are viewed as the walkers in CTQW. When all qubits are set on resonance with a uniform frequency $\omega$, the first term becomes a constant and can be neglected from the Hamiltonian. For the single-walker scenario mainly studied for CTQW-based algorithms~\cite{kadian2021quantum}, the on-site interaction $U_j$ vanishes and the dynamics of the whole qubit lattice is governed by the hopping terms  
\begin{equation}\label{reduced Hamiltonian}
H_{\rm QW}=\sum_{\langle j,k \rangle} g_{jk} (a^\dagger_j a_k + a_j a^\dagger_k).
\end{equation}

As the mapping is established, we discuss the direct connections between the SC qubit arrays and CTQWs on graphs. We define the following equivalences:

\begin{definition}{Vertex $\thicksim$ Node Qubit:}
Let $\mathcal{G}=(\mathcal{V},\mathcal{E})$ be a graph with $N$ vertices $\mathcal{V}$. Each vertex corresponds to a node qubit on the SC $(N+M)$-qubit array, where there are $M$ excess qubits.
\end{definition}

\begin{definition}{Edge $\thicksim$ Coupling:}
  An edge $e\in \mathcal{E}$ between any two vertices $j, k \in \mathcal{V}$ is equivalent to a nonzero coupling between corresponding qubits.    
\end{definition}

\begin{definition}{Hopping equivalence:}
Suppose there is an edge between $j$ and $k$ for some vertex $j, k \in \mathcal{G}$ then 

\begin{equation}
  \bra{k}A\ket{j}\equiv{\bra{b_k}} H_{\rm QW} \ket{b_j} = g_{jk}, 
\end{equation}
up to constant terms. Here, $H_{\rm QW}$ is the effective Hamiltonian
on the single-walker state space  spanned by the basis $\mathcal{B} = \{ | b_j \rangle = | 0 \rangle_1 | 0 \rangle_2 \ldots | 1 \rangle_j \ldots | 0 \rangle_N \mid 1 \leq j \leq N \}$. The matrix $[g_{jk}]$ constitutes the graph's adjacency matrix $A$, with $g_{jk}=J$ if vertices $j$ and $k$ are connected by an edge, and $g_{jk}=0$ otherwise. 
\end{definition}

In the context of quantum walks, our objective is to achieve uniform transition probabilities between graph vertices by equalizing the coupling strength among all vertex pairs and ensuring equal vertex frequencies. However, in superconducting qubits, residual couplings $g_{r}$ between remote qubits can be harmful to the effective CTQW model. Based on experimental data in Ref.\cite{guo2021stark,ni2022scalable}, the $g_r$ is typically $0.1$ to $1\ \text{MHz} \times 2\pi$. To ensure the efficacy of the simulated $H_{QW}$, the hopping strength $g_{jk}$ should be much greater than $g_r$. Hence, we set the lower limit $g_{jk}\geq3\ \text{MHz} \times 2\pi$. 

\begin{figure}[b]
    \includegraphics[width=0.85\linewidth]{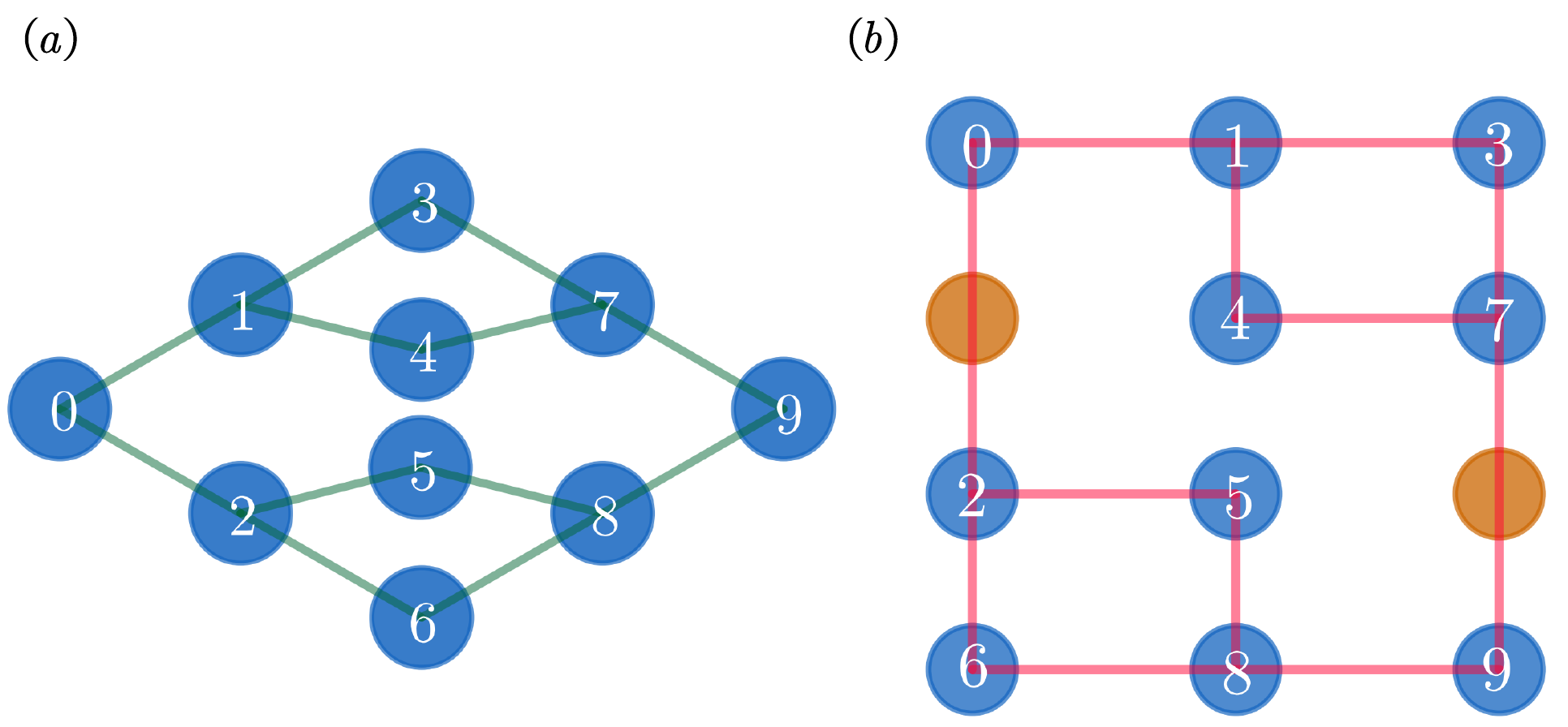}
    \caption{\textbf{Glued binary tree:} (a) A glued binary tree with layer three, which we aim to simulate on superconducting qubit arrays. (b) One possible implementation of (a) on a 3-by-4 qubit array using SEW. The \textcolor{blue}{blue} qubits are used as the graph vertices, while the \textcolor{orange}{orange} qubits are used to construct effective couplings between the blue vertices connecting to their two ends. 
    }
    \label{3treechip}
\end{figure}

Implementing complex graphs on simple low-dimensional qubit lattices confronts a critical challenge: Some graph vertices map to non-adjacent node qubits with excess qubits in between. For example, Fig.~\ref{3treechip} shows one way to map the glued binary tree to a 2D qubit lattice. The two excess qubits make it hard to establish an edge directly linking the non-adjacent qubits $\{0,2\}$ and $\{7,9\}$. The next two sections will present our solutions to weave edges using quantum bridges. As a result, complex graphs can be implemented on such qubit lattices. 

\section{Static Edges Weaving }

This section presents our first method, SEW, for enabling complex graph structures on low-dimensional qubit lattices. We will show that an effective edge between non-adjacent node qubits can be established by statically detuning the frequency of the connector qubits. We call the effective edges constructed in this way the static quantum bridge edge. The use of such bridge edges expands the lattice's connectivity and allows for the construction of diverse graph-based Hamiltonians, such as glued binary trees, complete graphs, and tetrahedron arrays, highlighting the method's utility in advancing quantum hardware design. 

\subsection{Detuned qubits as quantum bridge}\label{three_qubit}

\begin{figure}
    \centering
    \includegraphics[width = 0.85\linewidth]{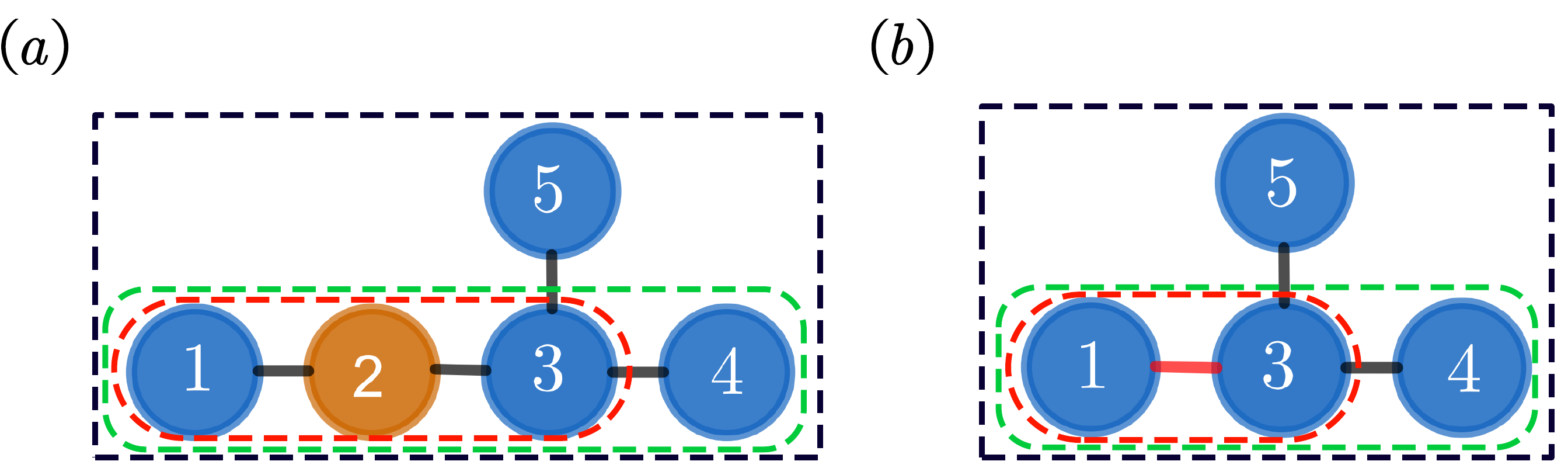}
    \caption{{\textbf{Static quantum bridge:}} (a) A physical implementation of the quantum bridge, where each nearest-neighbour pair of qubits is interconnected via a tunable coupler. While $Q_2$ is detuned appropriately, a bridge edge between $Q_1$ and $Q_3$ is effectively established as (b) shows. The qubits of interest are highlighted in \textcolor{blue}{blue}, whereas $Q_3$ used to facilitate effective coupling is marked in \textcolor{orange}{orange}. The dashed boxes of different colors represent the models we consider in the following section. (b) The effective model of (a), \textcolor{black}{black} line represents the direct connection, whereas the \textcolor{red}{red} line denotes the effective connection.}
    \label{SEWmodel}
\end{figure}

Consider a system of three transmon qubits~\cite{yan2018tunable} arranged in a chain, as depicted in the red dashed box of Fig.~\ref{SEWmodel}(a). The Hamiltonian governing this system is 
\begin{equation}\label{Eq_three_qubit_hamiltonian}
    \begin{aligned}
        H &= H_0+V, \\
        H_0 &= \sum_{i=1}^{3} \omega_{i} a_i^{\dagger} a_i +  \frac{\alpha_i}{2} a_i^{\dagger} a_i^{\dagger} a_i a_i, \\
        V &= g_{12}a_1^{\dagger}a_2 + g_{23}a_2^{\dagger} a_3+H.C.,
    \end{aligned}
\end{equation}
of where $\omega_i$ is the tunable transmon's frequency, $\alpha_i$ is the transmon's anharmonicity. Here, $H_0$ includes the three transmons' energies, while $V$ corresponds to their interaction. 

\begin{figure}[t]
    \centering
    \includegraphics[width =\linewidth]{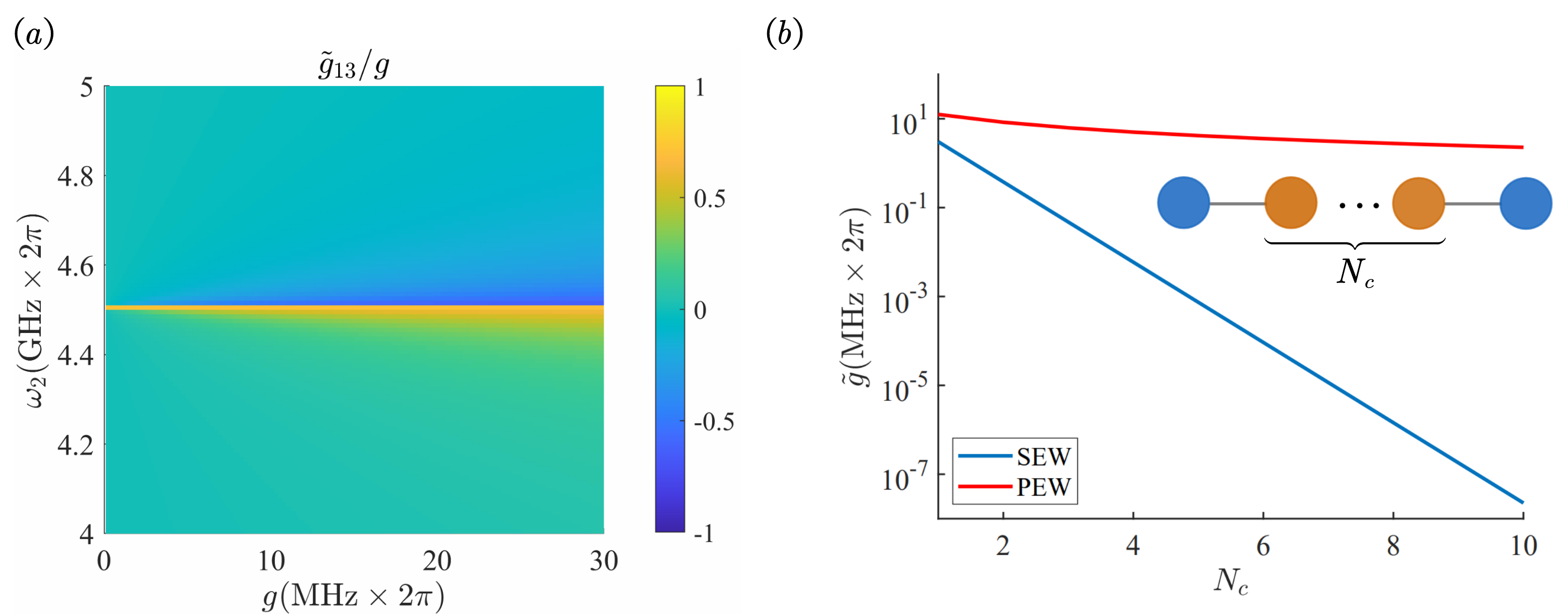}
    \caption{{\textbf{Effective coupling strength and scaling laws:}}
    (a) The numerical results of $\tilde{g}_{13}/g$ in a three-qubit chain (refer to Fig.~\ref{SEWmodel}(a)), obtained via the least action method. Here, $\tilde{g}_{13}$ is the effective hopping strength between $Q_1$ and $Q_3$, while $g$ is the direct coupling strength between neighbouring qubits. 
    (b) This figure illustrates that the effective coupling strength between the qubits at both ends decreases as the number of connector qubits $N_c$ increases. The \textcolor{blue}{blue} line and \textcolor{red}{red} line correspond to SWE and PEW, respectively. PEW will be introduced in the following section.
    }
    \label{qcq_hopping_scalling_law}
\end{figure}

By detuning $\omega_2$ away from $\omega_{1}$ and $\omega_3$, the effective Hamiltonian of this system could be derived using the Bloch perturbation theory~\cite{TAKAYANAGI2016200} (refer to Appendix~\ref{Appendix_H_eff_three_qubit_chain} for details). Therefore, the connector $Q_2$ now behaves as a static quantum bridge to establish an edge between $Q_1$ and $Q_3$ with the effective coupling strength up to the fourth order as
\begin{equation}\label{effective_coupling}
\begin{aligned}
     \Tilde{g}_{13} &= \frac{g_{12} g_{23} }{2} \Big[\frac{1}{\Delta_1}+\frac{1}{\Delta_3}-(\frac{g_{12}^2}{\Delta_1}+\frac{g_{23}^2}{\Delta_3})(\frac{1}{\Delta_1^2}+\frac{1}{\Delta_3^2}) \Big],
\end{aligned}
\end{equation}
where $\Delta_i=\omega_i-\omega_2, i\in \{1,3\}$. Note that Eq.~\ref{effective_coupling} is only valid under $g/\Delta_i\ll 1$.  To obtain this effective hopping more precisely, even beyond this limit, we apply an exact block-diagonalization numerical approach based on the least action principle (EBD-LA) given in ~\cite{Cederbaum1989BlockDO}. Solving the effective Hamiltonian on $Q_2$'s ground state subspace, we numerically solve the effective hopping strength $\tilde{g}_{13}$ as shown in Fig.~\ref{qcq_hopping_scalling_law}(a), with the dependency of $\tilde{g}_{13}/g$ on $\omega_2$ and $g$.   In our simulations, the transmons are truncated to three levels. We use realistic parameters for the numerical studies:  $\omega_1 = \omega_3 = 4.5\ \text{GHz}\times 2 \pi$; $\alpha_1=\alpha_2=\alpha_3 = -250\ \text{MHz}\times 2 \pi$; $g_{12}=g_{23}=g=25\ \text{MHz} \times 2\pi$;  We can see that $\tilde{g}_{13}$ is proportional to $g$ (when $\omega_2$ is away from resonance). It gradually increases as $\omega_2$ approaches to $4.5\ \text{GHz}\times 2 \pi$.  While fixing $\omega_2 =4.7\ \text{GHz}\times 2 \pi$ to suppress the excitation at $Q_2$, we obtain an effective edge with the coupling strength $\Tilde{g}_{13} \approx -3.1\ \text{MHz}\times 2\pi $. 

\begin{figure}[h]
    \centering
    \includegraphics[width = \linewidth]{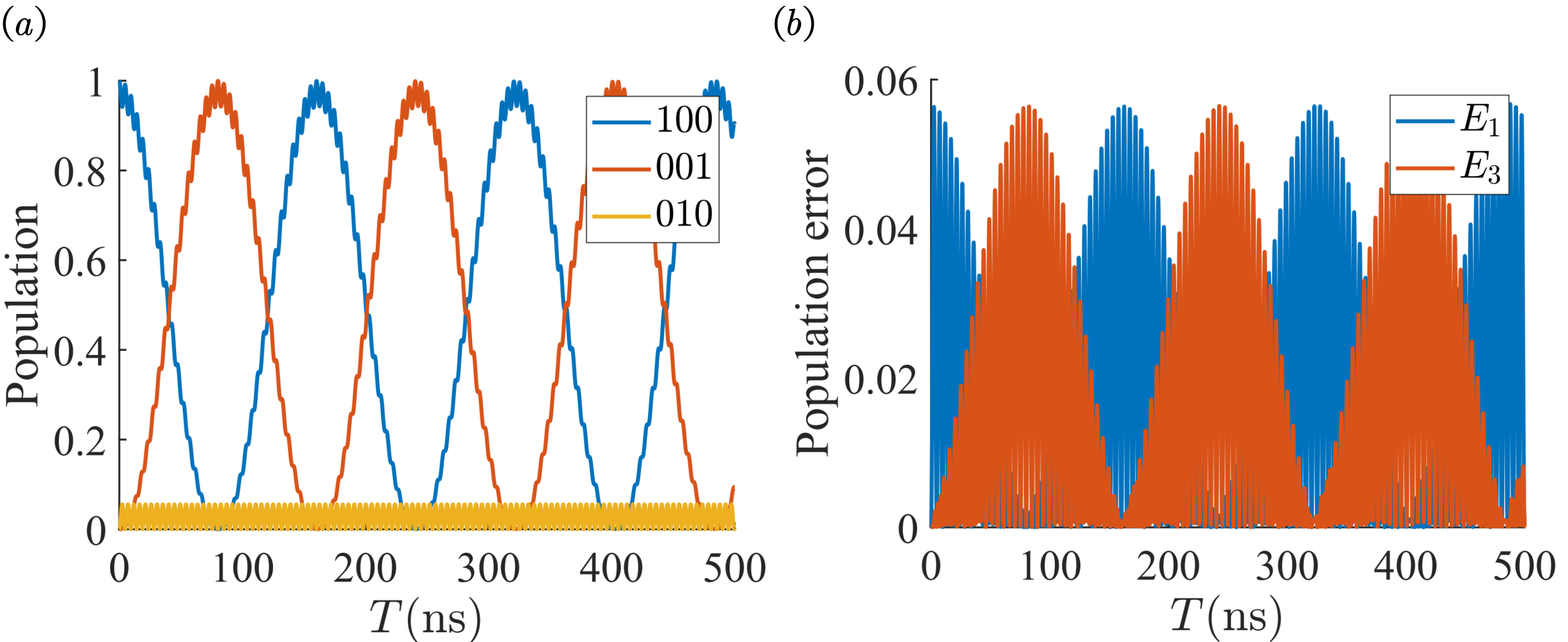}
    \caption{{\textbf{Population evolution of the three-qubit chain:}} (a) A rough oscillation between $Q_1$ and $Q_3$.  The initial state is $\ket{100}$, with detuning $\Delta_{12}=\Delta_{32}=-0.2\ \text{GHz}\times 2 \pi$ and direct coupling $g=25\  \text{MHz}\times 2 \pi$. (b) The population error of the three-qubit chain. The similar dynamics of both models validate the effectiveness of this scheme. Both simulations are done without additional counter-rotating terms in direct couplings.
    }
    \label{qcq_evol_error}
\end{figure}

The efficacy of this static bridge model could be further verified via the simulation of the system's dynamics. As shown in Fig.\ref{qcq_evol_error}, we plot the quantum states' evolution and its corresponding error $E_{k}(t):=\Big| \left|\bra{k} e^{-iHt} \ket{\varphi} \right|^2- \left|\bra{k} e^{-iHt} \ket{\varphi} \right|^2 \Big|$, which quantifies the population difference at $Q_k$  in time $t$ between the physical and effective models for a selected initial state $\ket{\varphi}$, $|.|$ is an absolute value. Here, $\ket{k}$ represents a state where the excitation is exclusively located on $Q_k$. This plot shows that the population on $Q_2$ is suppressed, and the small population errors on $Q_1$ and $Q_3$ indicate a consistent dynamic between the physical and effective model. 

Intuitively, a longer bridge slows down the walk speed. To quantitatively study this issue, the effective hopping strength is plotted in Fig.\ref{qcq_hopping_scalling_law}(b) to show the scaling versus the number of connectors $N_c$. The linear trend observed indicates that the SEW method has an exponential decrease law. Due to the limitation of a maximum of $g\geq3\ \text{MHz}\times 2 \pi$ introduced in the previous section, the length of the static quantum bridge is restricted to no longer than one connector.

\subsection{Expanding graphs with SEW}

In the following, we discuss how to weave complex graphs out of fundamental units: direct edges and bridge edges. Adding more elements to the lattice shifts the effective Hamiltonian. Verifying that the static bridge model holds for the expansion in different scenarios is important. 

\subsubsection{Expanding in 1D chain}
We first investigate whether the effective model of the static quantum bridge remains valid when expanding to a four-qubit one-dimensional chain. As shown in the green dashed box of Fig.~\ref{SEWmodel}, we use $Q_2$ as the connector to construct a bridge edge between $Q_1$ and $Q_3$ while linking $Q_3$ and $Q_4$ with a direct edge. By adding one qubit and one edge to the Eq.~\ref{Eq_three_qubit_hamiltonian}, we get the Hamiltonian of four-qubit chain.

Using the Bloch perturbation theory up to the fourth order, we get three effective hopping strengths:
\begin{equation}
\begin{aligned}
    \Tilde{g}_{13} &= \frac{g_{23}g_{12}}{2} \Big[\frac{1}{\Delta_3}+\frac{1}{\Delta_1} + \frac{g_{34}^2}{\Delta_4 \Delta_3^2} 
    - (\frac{g_{23}^2}{\Delta_3}+\frac{g_{12}^2}{\Delta_1})(\frac{1}{\Delta_3^2}+\frac{1}{\Delta_1^2}) \Big],\\
    \Tilde{g}_{34} &= g_{34}-\frac{g_{23}^2 g_{34}}{\Delta_3 \Delta_4}, \qquad
    \Tilde{g}_{14} =  -\frac{g_{34}g_{23}g_{12}}{\Delta_3 \Delta_4},
\end{aligned} 
\end{equation}
 where $\Delta_i = \omega_i-\omega_2,i\in \{1,3,4\}$. 

CTQWs require uniform walk speed (hopping rate) on each edge, which imposes a constraint on the effective hopping strength as $\tilde{g}_{13} = \tilde{g}_{34}=J$, where walk speed $J=3.1\ \text{MHz} \times 2\pi$ is the effective coupling strength of the static bridge obtained via Eq.~\ref{effective_coupling}. To verify this numerically, we use the parameter set $g_{34} =J= 3.1\ \text{MHz}\times 2\pi$, $g_{23} = g_{12} = 25\ \text{MHz} \times 2\pi$, and $\Delta_1 = \Delta_3 = \Delta_4 = -200\ \text{MHz} \times 2\pi $. We obtain $\tilde{g}_{13} \approx 3.17\ \text{MHz}\times 2\pi$, with the relative error $|(\tilde{g}_{24}-J)/J|\approx 2.2\%$. This implies that the impact on the effective hopping between $Q_1$ and $Q_3$ is small enough by adding $Q_4$, thereby confirming the Extensibility of the static quantum bridge model and, hence, the SEW method. 

\begin{figure}[h]
    \centering
    \includegraphics[width = \linewidth]{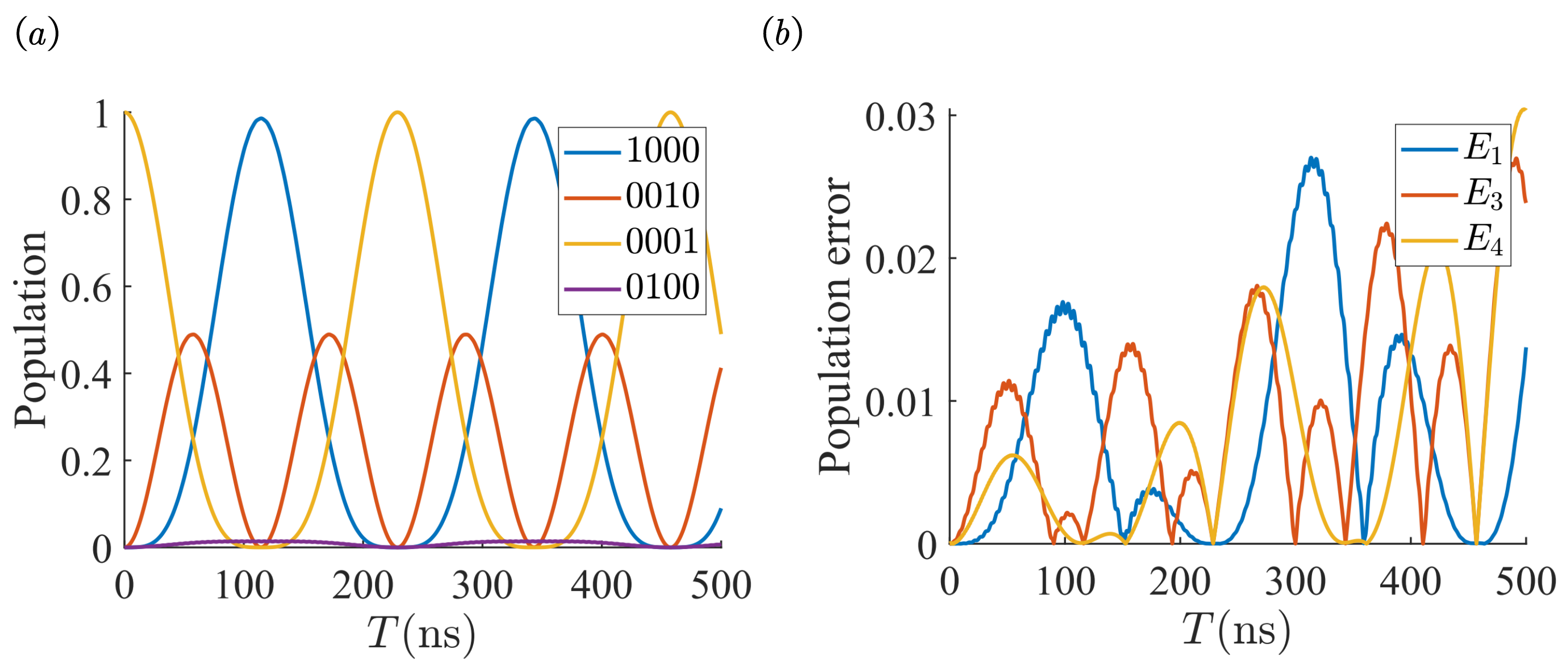}
    \caption{{\textbf{Population evolution of the four-qubit chain:}} (a) Populations for each qubit in the physical model, which is shown in the \textcolor{green}{green} dashed box of Fig.~\ref{SEWmodel}(a). The initial state is $\ket{0001}$. (b) The population errors for each \textcolor{blue}{blue} qubit. The small errors certify the similar dynamics of the two models and the validity of the effective model. }
   \label{qqcq_evol_error}
\end{figure}

Due to the frequency shift on $Q_1$ and $Q_3$, we slightly adjust the frequency $\omega_4 = 4.497\ \text{GHz}\times 2\pi$ to ensure that $\tilde{\omega}_1=\tilde{\omega}_3=\tilde{\omega}_4$.
Like the last section, we plot the evolution of the four-qubit chain and the three population errors $E_1, E_3, E_4$ for the three node qubits in Fig.~\ref{qqcq_evol_error}. As shown in Fig.~\ref{qqcq_evol_error}(a), the excitation on the coupler $Q_2$ is strongly suppressed under the parameters. The small population errors presented in Fig.~\ref{qqcq_evol_error}(b) verify the feasibility of the effective model for the four-qubit chain.

\subsubsection{Expanding to 2D chain}
Realistic qubit lattices are usually two-dimensional. Here, we investigate our model's extensibility to a two-dimensional lattice. The specific lattice, depicted in the black dashed box of Fig.~\ref{SEWmodel}(a), consists of two qubits connected directly to the $Q_3$, while the third qubit is indirectly connected. Notably, $Q_3$ resides on the graph's edge, leading to the effective model shown in Fig.~\ref{SEWmodel}(b). Following the same workflow as the previous discussion, taking $\Delta_i = \omega_i-\omega_2,\ \Sigma_{13} = g_{23}^2/\Delta_3+g_{12}^2/\Delta_1$, $\Sigma_{45} = g_{34}^2/\Delta_4+g_{35}^2/\Delta_5$, 

all the effective hopping strength could be derived as: 

\begin{equation}
    \begin{aligned}
        \Tilde{g}_{13} &= \frac{g_{23} g_{12}}{2} \left(\frac{1}{\Delta_3} + \frac{1}{\Delta_1} + \frac{\Sigma_{45}}{\Delta_3^2} - (\frac{1}{\Delta_3^2}+\frac{1}{\Delta_1^2})\Sigma_{13}\right),\\
        \Tilde{g}_{34} &= g_{34}-\frac{g_{34}g_{23}^2}{\Delta_3 \Delta_4} , \qquad
        \Tilde{g}_{35} = g_{35} -\frac{g_{35}g_{23}^2}{\Delta_3 \Delta_5} ,\\
        \Tilde{g}_{14} &= -\frac{g_{{12}}g_{{23}}g_{{34}}}{\Delta_3 \Delta_4},\qquad
        \Tilde{g}_{{15}} = -\frac{g_{{35}}g_{{23}}g_{{12}}}{\Delta_{3} \Delta_{5}}.\\
    \end{aligned}
\end{equation}

\begin{figure}[b]
    \centering
    \includegraphics[width = \linewidth]{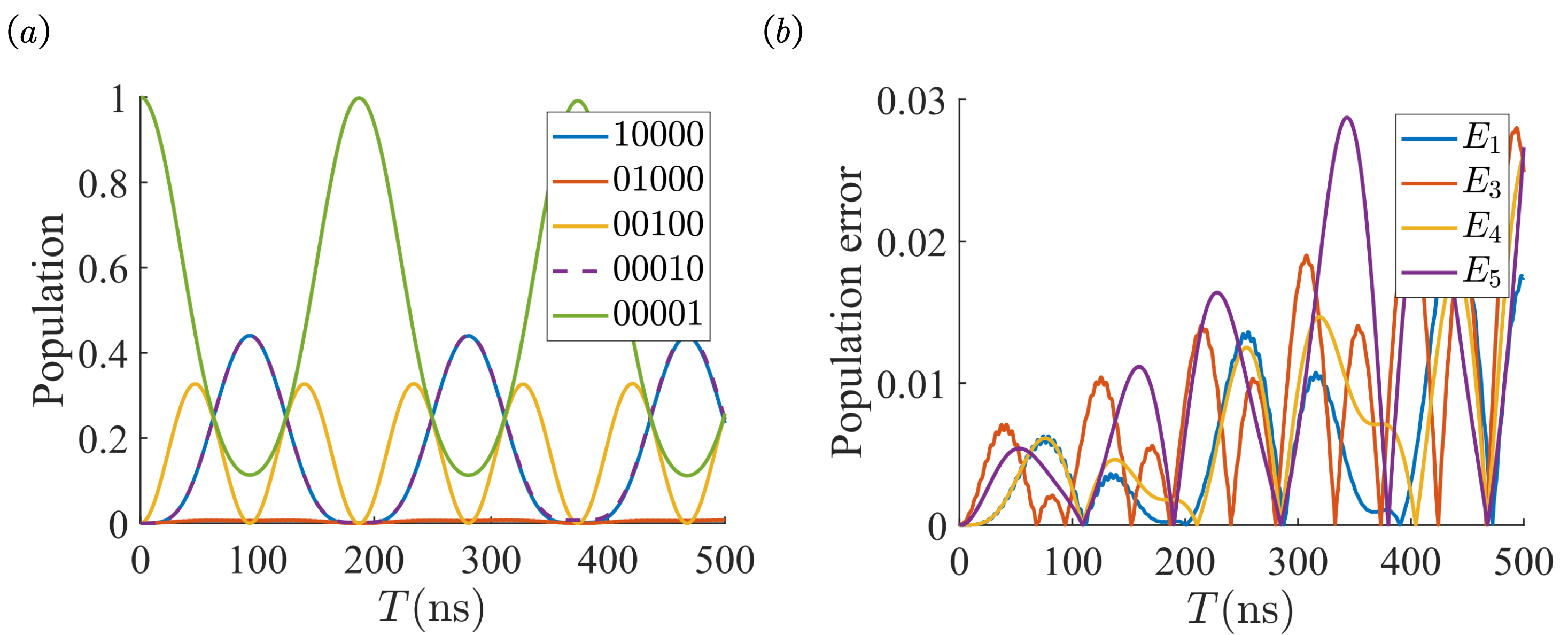}
    \caption{{\textbf{Population evolution of the 2D model:}} (a) Populations for each qubit in the physical model, which is shown in the black dashed box of Fig.~\ref{SEWmodel}(a). The initial state is $\ket{10000}$. (b) The population errors for each \textcolor{blue}{blue} qubit.}
    \label{qqqcq_evol_error}
\end{figure}

To satisfy the uniform walk speed constraint $\tilde{g}_{13} = \tilde{g}_{34} = \tilde{g}_{35}=J$, we set the following parameters for the numerical study of this case: the direct connections $g_{34} = g_{35} = J=3.1\ \text{MHz}\times 2\pi$, and $g_{23} = g_{12} = 25\ \text{MHz}\times 2\pi$, $\Delta_{1}=\Delta_{3}=-200 \text{ MHz}\times 2\pi$, $\Delta_4 = \Delta_5 = -203 \text{MHz} \times 2\pi$. Similar to the 1D chain, we increase the frequencies of qubits connected to $Q_{2}$, and the coupling strength between them is fine-tuned to reduce the error. We then obtain the effective hopping strength $\tilde{g}_{13}\approx -3.16\text{ MHz}\times 2\pi$ for the bridge edge with a relative error of  $\approx0.19\%$. Still, the impact remains small enough by adding $Q_5$ and expanding to a 2-dimensional structure. This could be further verified via the numerical simulation of the dynamics. As depicted in Fig.~\ref{qqqcq_evol_error}, the population error remains sufficiently small, where the slow increasing error over time could be suppressed by further optimizing the parameters.

\begin{figure}[b]
    \centering
    \includegraphics[width=\linewidth]{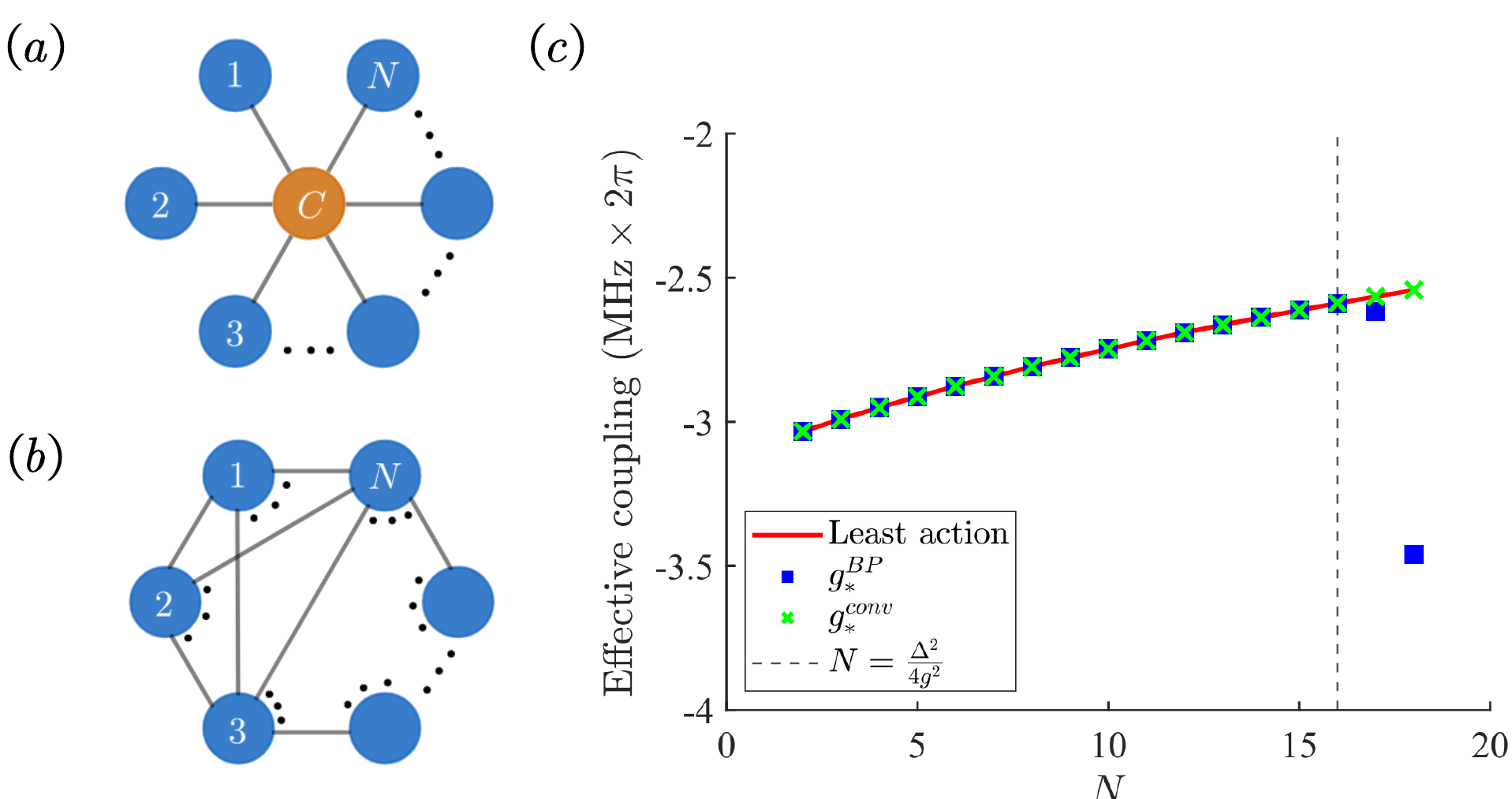}
    \caption{{\textbf{Star topology:}} (a) The star topology consists of $N$ peripheral qubits uniformly coupled to a central hub qubit. Each peripheral qubit exhibits identical coupling strength to the hub. All peripheral qubits (depicted in \textcolor{blue}{blue}) share equivalent frequency and anharmonicity parameters, while the central qubit (depicted in \textcolor{orange}{orange}) facilitates the construction of effective couplings. (b) The corresponding effective model represents a fully connected graph with $N$ vertices. (c) Comparison of perturbation theory and the least action method. The \textcolor{red}{red} line represents the effective couplings calculated using the least action method, while the \textcolor{green}{green} cross illustrates the theoretical limit of perturbation theory as per Eq.~\ref{star_conv_coupling}. The \textcolor{blue}{blue} square shows perturbation theory predictions up to the first 100 orders (truncated at $p=50$ in Eq.~\ref{star_pt}). The \textcolor{black}{black} vertical dashed line marks the threshold value of $N$ beyond which perturbation theory ceases to be valid.}
    \label{star_model_num}
\end{figure}
\subsubsection{Expanding to higher dimensions with star graphs}

In pursuing scalable quantum computing architectures, the complete connectivity as a higher dimensional topology emerges as a pivotal structure, which can be constructed using a 2D star graph (depicted in Fig.~\ref{star_model_num}(a)) with the static quantum bridge in the center. The Hamiltonian of this system is given by:
\begin{equation}
\begin{aligned}
        H &= H_{0} + V,\\
        H_0 &= \omega_{c} a_{c}^{\dagger} a_{c} + \frac{\alpha_{c}}{2} a_{c}^{\dagger} a_{c}^{\dagger} a_{c} a_{c} + \sum_{i=1}^{N} (\omega_{i} a_{i}^{\dagger} a_{i} + \frac{\alpha_{i}}{2} a_{i}^{\dagger} a_{i}^{\dagger} a_{i} a_{i}),\\
    V&= \sum_{i=1}^{N} g_{i} a_{i}^{\dagger} a_{c} +H.C.,
\end{aligned}
\end{equation}
where $\omega_{c}$ and $\omega_{i}$ represent the frequencies of the central and peripheral qubits, respectively, $\alpha_{c}$ and $\alpha_{i}$ their anharmonicities, and $g_{i}$ the coupling strength between the $i$-th peripheral qubit and the central qubit.

The effective coupling between any two peripheral qubits, assuming uniform frequency $\omega$ and coupling strength $g$, can be derived using Bloch perturbation theory as:
\begin{equation}\label{star_pt}
    \tilde{g}_{*}^{BP} = \frac{g^2}{\Delta} \sum_{p=0}^{\infty} C_p \left(-N \frac{g^2}{\Delta^2}\right)^p,
\end{equation}
where $\Delta = \omega - \omega_c$ and $C_p$ are the Catalan numbers. 
The series converges when $N < \frac{\Delta^2}{4g^2}$, as an extra constraint along with the perturbative condition. Then, the effective coupling is succinctly expressed as:
\begin{equation}\label{star_conv_coupling}
    \tilde{g}_{*}^{conv} = \frac{\Delta}{|\Delta|} \frac{\sqrt{\Delta^2 + 4Ng^2} - |\Delta|}{2N}.
\end{equation}

We again use the numerical approach EBD-LA~\cite{Cederbaum1989BlockDO} to verify the effective couplings derived above. As illustrated in Fig.~\ref{star_model_num}(c), the theoretical upper limit for $N$ is determined to be $\frac{\Delta^2}{4g^2}$. It is observed that the perturbative outcomes begin to diverge from those predicted by the least action principle beyond a specific $N$ threshold. Intriguingly, for $N \geq \frac{\Delta^2}{4g^2}$, Eq. ~\ref{star_conv_coupling} aligns with the least action principle results, as evidenced in Fig.~\ref{star_model_num}(c). Even though we do not have an explanation for this, the formula can be directly utilized to compute the effective coupling strength. Let us consider a system with $N = 3$ qubits, where the detuning is $\Delta = -200\text{ MHz}\times 2\pi$, the coupling strength is $g = 25\text{ MHz}\times 2\pi$ and the anharmonicity is uniformly $-250\text{ MHz}\times 2\pi$. In this case, the calculated effective coupling strength is approximately $-2.99\text{ MHz}\times 2\pi$.

As Fig.~\ref{star_model_num}(b) shows, our examination of the star topology reveals its equivalence to a complete graph characterized by uniform effective couplings among peripheral qubits. This uniformity in effective coupling is a hallmark of complete graphs. 

\subsection{Weaving Arbitrary Graphs with Static Bridges}

Previously, we quantitatively verified three fundamental structures to weave the graph from one dimension to higher dimensions by combining static quantum bridges with direct edges. These structures enable excess possibilities in creating complex graph configurations on basic lattice frameworks. Here we explicitly present the construction of a nontrivial graph, the glued tetrahedron. 

 \begin{figure}[b]
    \centering
    \includegraphics[width=0.85\linewidth]{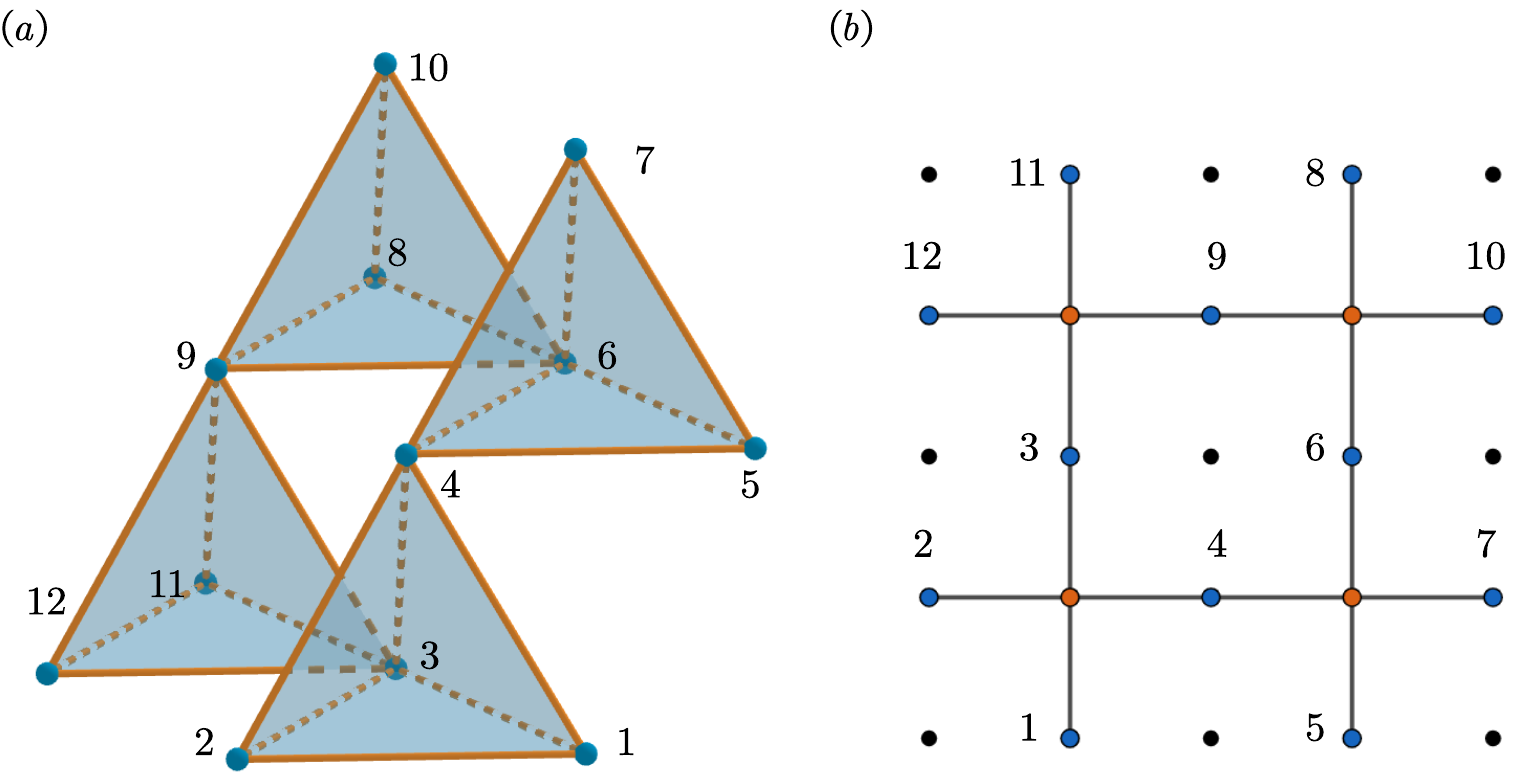}
    \caption{{\textbf{Glued tetrahedron array and its layout method:}} (a) The 3D graph is constructed using tetrahedron graphs, representing the vertices of the graph. (b) On the chip, we implement the graph using qubits. The numbers correspond to the same vertices as in (a). The \textcolor{blue}{blue} points represent the qubits used to implement the vertices, 
    while the \textcolor{orange}{orange} points represent the qubits used to create effective couplings. 
    The black points indicate the unused qubits on the chip. The black line represents the connection scheme. 
    } 
    \label{Tetrahedron}
\end{figure}

Implementing three-dimensional (3D) graphs on a two-dimensional (2D) qubit lattice presents significant challenges. However, by employing the coupler technique, we have successfully demonstrated the construction of a 3D graph. As depicted in Fig.~\ref{Tetrahedron}(a), we achieve this by stacking tetrahedron arrays, each sharing vertices with its neighbours, to form a ladder-like structure. This method allows for the extension of the graph into the third dimension, with the potential to become a regular graph as the number of tetrahedrons increases indefinitely.

\section{Periodic Edges Weaving}

    In this section, we present our second method, periodic edge weaving (PEW), to obtain effective quantum walks between qubits at even longer distances, which we call dynamic quantum bridges. This helps overcome SEW's limitation in implementing complex graph structures, such as Hypercubes or fullerene graphs (see Fig.~\ref{fullerene_implement}). 
    
\subsection{Floquet graph engineering}

We utilize periodic engineering of dynamic graphs by switching on and off the tunable couplers between qubits. By carefully designing and repeating dynamic graph sequences, we construct a Floquet Hamiltonian~\cite{montagnier2004control}
that facilitates effective quantum walks on the expected graphs. 

\begin{figure}[h]
    \centering
    \includegraphics[width = \linewidth]{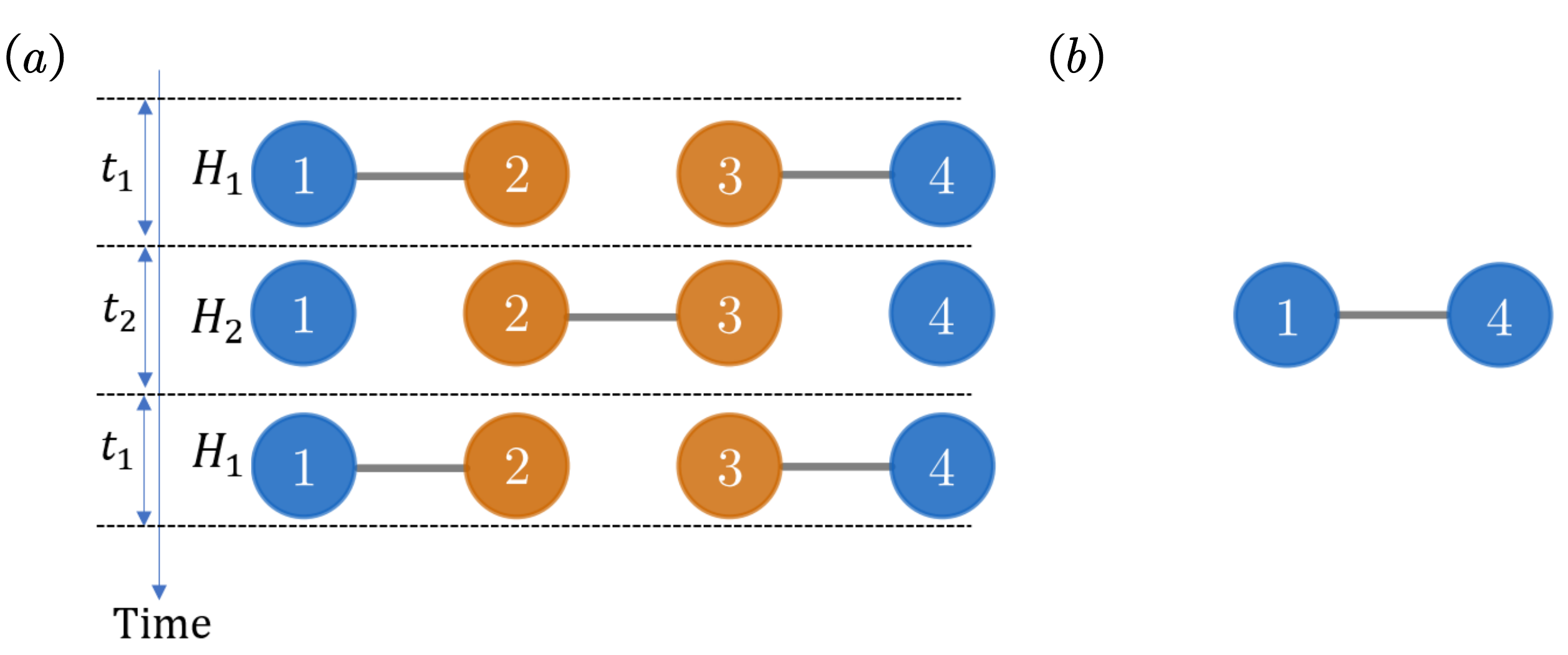}
    \caption{{\textbf{Dynamic graph in one period and scaling law of two methods:}} (a) The qubits depicted in \textcolor{blue}{blue} represent the vertices used in the graph, while the \textcolor{orange}{orange} qubits are connectors to construct effective couplings. $t_1$ and $t_2$ denote the time intervals during which the corresponding Hamiltonians, $H_1$ and $H_2$, are applied. (b) The figure illustrates an effective model obtained from (a) at a specific time, which is the integer multiple of the period.}
    \label{q4d_model}
\end{figure}

Consider a qubit array $Q_1, Q_2, Q_3, Q_4$, the Floquet control involves a sequence of periods, as shown in Fig.~\ref{q4d_model}(a). During time $t_1$, the coupling $g_{12}$ and $g_{34}$ are turned on while $g_{23}=0$. In the subsequent time period $t_2$, the coupling $g_{23}$ is on while $g_{12}=g_{34}=0$, followed by a repetition of $t_1$. The effective model of this Floquet system is shown in Fig.~\ref{q4d_model}(b) where $Q_1$ and $Q_4$ are connected with $Q_2, Q_3$ together as the quantum bridge. The quantum walk dynamics between $Q_1$ and $Q_4$ are mimicked in the effective model when measurements are performed at the end of a Floquet period. We analyze the Hamiltonians for control sequence $t_1$ and $t_2$, denoted as $H_1$ and $H_2$, focusing on the one-excited subspace. Here, the Hamiltonians are expressed as $H_i = \omega I + g A_i$, where $I$ represent the four-dimensional identity matrix and $A_i$ is the adjacency matrix for $i \in \{1,2\}$, given by Eq.~\ref{adjacency}

By setting $t_1 = \frac{\pi}{2g}$, the effective coupling strength between $Q_1$ and $Q_4$ at the total time $2t_1+t_2$ is derived as (refer to Appendix~\ref{Appendix_H_eff_floquet})

$$
\tilde{g} = g \frac{g t_2}{g t_2+\pi}.
$$
This effective hopping strength can be verified numerically by simulating the quantum walk between $Q_1$ and $Q_4$, as discussed in the next subsection. 

\subsection{Effective quantum walk on Floquet graphs}

\begin{figure}[h]
    \centering
    \includegraphics[width = 0.9\linewidth]{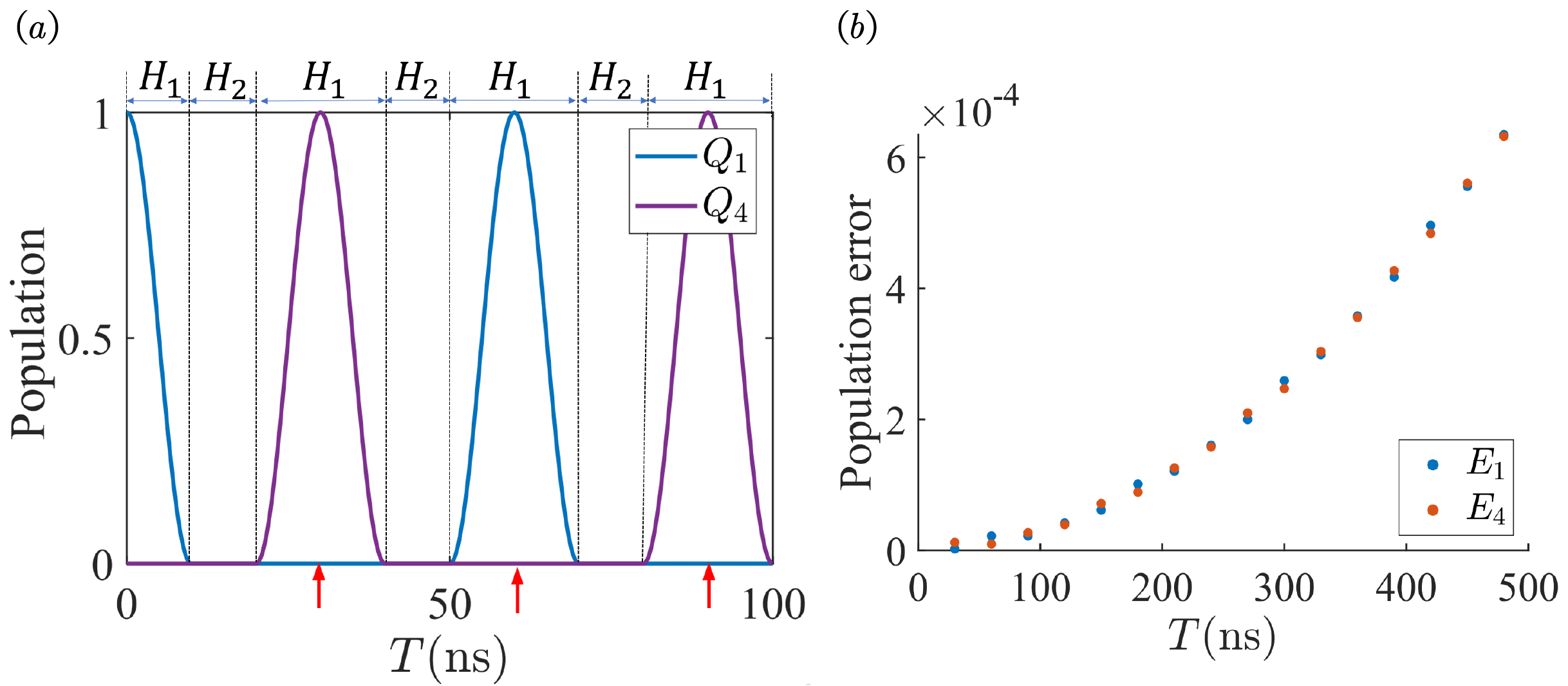}
    \caption{{\textbf{Population evolution of four-qubit dynamic control:}} (a) The population of $Q_1$ and $Q_4$ over time. The top axis represents the corresponding Hamiltonian of the system during each control sequence, and the \textcolor{red}{red} arrow indicates the time at which measurements are performed. (b) The error is defined as the difference between the model presented in Fig.~\ref{q4d_model} (a) and Fig.~\ref{q4d_model}(b) when measurements are taken. The figure demonstrates that the difference between the models is sufficiently small. 
    The physical model parameters used are $\omega_1 = \omega_2 = \omega_3 = \omega_4 = 4.5\ \text{GHz}\times 2 \pi$  and $g_{12} = g_{23} = g_{34} = 25\ \text{MHz}\times 2\pi$, while the effective model parameters are $\tilde{\omega}_1 = \tilde{\omega}_4 = 4.5\ \text{GHz} \times 2 \pi$ and $\tilde{g}_{14} = 25/3\ \text{MHz} \times 2\pi$.}
    \label{q4d_evol_error}
\end{figure}

For Floquet graphs, the effective quantum walks depend on the precise control of coupling times. By setting $t_2 = t_1 = \frac{\pi}{2g}$, we can suppress the population leakage to couplers $Q_2$ and $Q_3$ to guarantee the effective model.
We run the numerical simulation for the full Hamiltonian of the four-qubit array as Fig.~\ref{q4d_model}(a) shows. The observed population and error are presented in Fig.~\ref{q4d_evol_error}. The negligible magnitude of this error confirms the efficacy of the dynamic graph approach.

While utilizing multiple qubits as the dynamic quantum bridge is advantageous, we observed that an increase in connector qubit counts inversely affects the long-range coupling strength. Floquet control, characterized by its temporal precision, permits the establishment of effective long-range couplings with a minimal qubit count. For two neighboring qubits with a coupling strength $g$, the evolution time is given by $t=\frac{1}{4g}$, which is a quarter of the Rabi oscillation period. Following the same weaving patterns, with $N_c$ connectors to build the bridge, there are $N_c+1$ control fragments of each period, and the total time taken is $t_N=\frac{N_c+1}{2g}$. It counts for half of Rabi oscillation in the effective model, resulting in the equation $\frac{1}{2\tilde{g}} = \frac{N_c+1}{2g}$,  where $\tilde{g}$ is the effective coupling strength and 
\begin{equation}
    \tilde{g}=\frac{g}{N_c+1}.
\end{equation}
As shown in Fig.~\ref{qcq_hopping_scalling_law}(b), such a scaling law (dissipation of $\tilde{g}$ with an increase in $N_c$) is much better than the exponential decay of the SEW method. For example, to achieve an effective coupling strength of $3 \times 2\pi$ MHz, the system can accommodate up to $N_c=7$ qubits.

\section{Complex Graph Implementation via PEW}

\begin{figure}[h]
    \centering
    \includegraphics[width=0.85\linewidth]{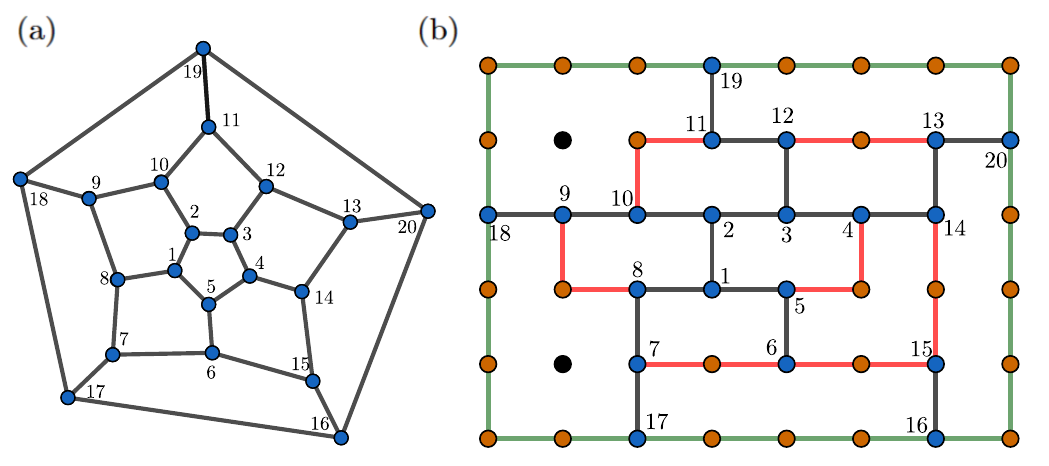}
    \caption{{\textbf{Fullerene and its layout method:}} (a) Fullerene  20 Schlegel diagram. (b) Implementation of the Fullerene graph on a grid lattice. \textcolor{blue}{Blue} vertices represent vertices from the Fullerene graph, while \textcolor{orange}{orange} vertices represent qubits used to construct effective couplings. Black lines denote static couplings, while \textcolor{green}{green} lines represent dynamic couplings.}
    \label{fullerene_implement}
\end{figure}

The PEW method obtains much longer quantum bridges and, hence, expands the class of graphs that can be realized beyond the qubit lattice's intrinsic connectivity and SEW's connectivity. This section illustrates the application of this technique to an exemplar graph: the fullerene 20, which cannot be implemented directly on a 2D qubit lattice or using the SEW method. It is chosen for specific reasons. Extensive research has been conducted to investigate faster quantum transport in graphene-like structures, particularly fullerenes, which have demonstrated promising transport properties~\cite{bougroura2016quantum}. These studies contribute to our understanding of the dynamic interplay between classical and quantum behaviour within general structures~\cite{rebentrost2009environment}. The fullerene-20 graph is shown in Fig.~\ref{fullerene_implement}(a). We detail a lattice-based methodology for constructing this graph, using a combination of SEW and PEW, as demonstrated in  Fig.~\ref{fullerene_implement}(b). While this approach necessitates feasibility verification and potential coupling adjustments, it holds promise for the realization of complex graph structures through the synergistic use of static and dynamic controls.

\section{Conclusion}

In this work, we have presented two novel approaches, static and periodic edge weaving, to overcome the limited qubit connectivity inherent in low-dimensional qubit arrays, with a particular focus on superconducting qubit chips. The SEW approach utilizes detuned qubits as effective quantum bridges to link non-adjacent vertices, which realizes diverse graph-based Hamiltonians on planar lattices, such as binary tree graphs, complete graphs and glued tetrahedrons. The PEW approach uses Floquet control of qubit couplings (the edges) to facilitate effective quantum walks with extended-ranged vertices, granting access to intricate graphs like cubes and fullerenes. Numerical evidence confirms the fidelity of this technique in transcending the inherent connectivity constraints. The synergistic combination of these two methodologies unlocks a vast design space for realizing complex quantum networks tailored for diverse computational tasks. 

The scaling law of the effective walk speed over the length of bridges is studied. The effective Hamiltonian’s coupling strength diminishes as the number of qubits in the coupler chain increases. For SEW, the effective coupling strength decays exponentially, as indicated by numerical results, whereas, for PEW, the analytical result reveals it scales as $1/N_c$. Consequently, the PEW exhibits a more favourable scaling law, albeit at the expense of fixed measurement periods, in contrast to the static detuning approach, which permits measurements at any time.

In summary, our work paves the way for maximizing the capabilities of state-of-the-art quantum hardware, pushing the boundaries of analog quantum simulation, particularly in the realm of continuous-time quantum walks. Future research will focus on the scalability and fidelity of the combination of the two approaches, their integration into a unified model, and their application to various quantum platforms. 




\appendices
\section{Effective coupling on three-qubit chain} \label{Appendix_H_eff_three_qubit_chain}

We utilize Bloch perturbation theory~\cite{TAKAYANAGI2016200} to derive the effective Hamiltonian. Starting from the Hamiltonian described in Eq.~\ref{Eq_three_qubit_hamiltonian}, we focus on simplifying the analysis by considering the subspace composed of a single excited state. Assume $P=\ket{100}\bra{100}+\ket{001}\bra{001}$ is the projection operator for subspace spanned by $S_P = \{\ket{100}, \ket{001} \}$, and $Q=\ket{010}\bra{010}$ is the projection operator for subspace spanned by $S_Q = \{\ket{010}\}$. The effective Hamiltonian is 
\begin{equation}
   \label{bloch_perturbation}
    \begin{aligned} 
        H_{\rm{eff}} &= PH_0 P + V_{\rm{eff}},\\
    \end{aligned}
\end{equation}
where $V_{\rm{eff}}$ is effective coupling. 

 We get the effective Hamiltonian of Eq.~\ref{Eq_three_qubit_hamiltonian} with truncation up to the fourth order as follows
\begin{equation}
    \begin{aligned}
        V_{\rm{eff}}^{(1)}=& 0,\\
        V_{\rm{eff}}^{(2)}  =& \frac{g_{12} g_{23}}{\Delta_1} \ket{001}\bra{100}+\frac{g_{12}^2}{\Delta_1} \ket{100}\bra{100}\\
            &+\frac{g_{12} g_{23}}{\Delta_3}\ket{100}\bra{001}+\frac{g_{23}^2}{\Delta_2} \ket{001}\bra{001},\\
        V_{\rm{eff}}^{(3)}  =& 0,\\
        V_{\rm{eff}}^{(4)}  =& -\frac{g_{12} g_{23}}{\Delta_1^2} \left( \frac{g_{12}^2}{\Delta_1} + \frac{g_{23}^2}{\Delta_3}\right)\ket{001}\bra{100}\\
        &-\frac{g_{12}^2}{\Delta_1^2}\left(\frac{g_{12}^2}{\Delta_1} + \frac{g_{23}^2}{\Delta_3} \right)\ket{100}\bra{100}\\
         & - \frac{g_{12}g_{23}}{\Delta_3^2}\left(\frac{g_{12}^2}{\Delta_1}+\frac{g_{23}^2}{\Delta_3} \right) \ket{100} \bra{001}  \\
         &- \frac{g_{23}^2}{\Delta_3^2}\left(\frac{g_{12}^2}{\Delta_1}+\frac{g_{23}^2}{\Delta_3} \right)\ket{001}\bra{001},\\ 
        \end{aligned}
\end{equation}
where $\Delta_i = \omega_i-\omega_2$,  for $i \in \{1,3\} $.

The effective coupling strength between $Q_1$ and $Q_3$ corresponds to the coefficient of $\ket{001}\bra{100}$ and $\ket{100}\bra{001}$ in the effective Hamiltonian. It is important to note that the effective Hamiltonian is non-Hermitian, resulting in different strengths for these two terms. To obtain a meaningful measure of the effective coupling between $Q_1$ and $Q_3$, we average their strengths. This approach addresses the non-Hermitian nature of the effective Hamiltonian within the framework of Bloch perturbation theory. Finally, the effective coupling strength between $Q_1$ and $Q_3$ is given by 

\begin{equation}
    \Tilde{g}_{13} = \frac{g_{12} g_{23} }{2} \Big[\frac{1}{\Delta_1}+\frac{1}{\Delta_3}-(\frac{g_{12}^2}{\Delta_1}+\frac{g_{23}^2}{\Delta_3})(\frac{1}{\Delta_1^2}+\frac{1}{\Delta_3^2}) \Big].
\end{equation}

Note that the resulting effective Hamiltonian includes additional terms, namely $\ket{001}\bra{001}$ and $\ket{100}\bra{100}$. These terms contribute to the frequency shift on $Q_1$ and $Q_3$.

\section{Effective Hamiltonian of Floquet graph}
\label{Appendix_H_eff_floquet}
As shown in Fig.~\ref{q4d_model}, the Hamiltonian within period $t_1$ is $H_1$ and period $t_2$ is $H_2$. We focus on the one-excited subspace, the Hamiltonian become $H = \omega I + g A_i$, where $I$ is the four dimension identity matrix and $A_i$ is the adjacency matrix given by
\begin{equation}\label{adjacency}
    A_1 = \begin{pmatrix}
         0 &1 &0 &0  \\
         1 &0 &0 &0  \\
         0 &0 &0 &1  \\
         0 &0 &1 &0  \\
    \end{pmatrix},
    A_2 = \begin{pmatrix}
         0 &0 &0 &0  \\
         0 &0 &1 &0  \\
         0 &1 &0 &0  \\
         0 &0 &0 &0 \\
    \end{pmatrix}.
\end{equation}
Then the evolution unitary of one period is 
\begin{equation}
\begin{aligned}
    U &= e^{-i H_1 t_1} e^{-i H_2 t_2} e^{-i H_1 t_1} \\
      &= e^{-i (\omega I+g A_1) t_1} e^{-i (\omega I+g A_2) t_2} e^{-i (\omega I+g A_1) t_1}\\
      &= e^{-i \omega (2t_1 + t_2) I} e^{-i A_1 g t_1} e^{-i A_2 g t_2} e^{-i A_1 g t_1}.\\
\end{aligned}
\end{equation}
We find when $t_1 = \frac{\pi}{2g}$, the unitary have a simple form
\begin{equation}\label{Ueff}
\begin{aligned}
        U &= - e^{-i \omega (2t_1 + t_2)I} \begin{pmatrix}
        \cos{(g t_2)} &0 &0 &-i\sin{(gt_2)}\\
        0 &1 &0 & 0\\
        0 &0 &1 &0\\
        -i\sin{(g t_2)} &0 &0 &\cos{(g t_2)}\\
    \end{pmatrix}\\
          & = - e^{-i \omega (2t_1 + t_2)I} e^{-i A_{\rm{eff}} g t_2} \\
          & = -e^{-i \Big(\omega I + g\frac{g t_2}{g t_2+\pi} A_{\rm{eff}}\Big)(2t_1+t_2)},
\end{aligned}
\end{equation}
where 
\[A_{\rm{eff}} = \begin{pmatrix}
0 &0 &0 &1\\
0 &0 &0 &0\\
0 &0 &0 &0\\
1 &0 &0 &0\\
\end{pmatrix}.\]

The global phase in Eq.~\ref{Ueff} can be ignored. Thus we get the effective Hamiltonian of this Floquet system as follows: 
\begin{equation}
    H_{\rm{eff}} = \omega I + g\frac{g t_2}{g t_2+\pi} A_{\rm{eff}} ,
\end{equation}
with effective coupling strength between $Q_1$ and $Q_4$
\begin{equation}
    \Tilde{g} = g \frac{g t_2}{g t_2+\pi}.
\end{equation}

\end{document}